\documentstyle[referee]{mn}
\def\ut#1{\mathop{\vtop{\ialign{##\crcr
     $\hfil\displaystyle{#1}\hfil$\crcr\noalign
     {\kern1pt\nointerlineskip}\hbox{$\hfil\sim\hfil$}\crcr
     \noalign{\kern1pt}}}}}

\def\undersymbol#1#2{\mathop{\vtop{\ialign{##\crcr
     $\hfil\displaystyle{#2}\hfil$\crcr\noalign
     {\kern1pt\nointerlineskip}\hbox{$\hfil#1\hfil$}\crcr
     \noalign{\kern1pt}}}}}
\def\arcsec{^{\prime\prime}}

\begin{document}

\title[BINARY BROWN DWARFS IN THE GALACTIC HALO?]
      {BINARY BROWN DWARFS IN THE GALACTIC HALO?}

\author[F. De Paolis, G. Ingrosso, Ph. Jetzer and M. Roncadelli] 
{F.~De Paolis$^1$, G.~Ingrosso$^2$ , Ph.~Jetzer$^1$
and M.~Roncadelli$^3$
\\
$^1$
Paul Scherrer Institute, Laboratory for Astrophysics, CH-5232 Villigen
PSI, 
and\\
Institute of Theoretical Physics, University of Zurich, Winterthurerstrasse
190, CH-8057 Zurich, Switzerland\\
$^2$
Dipartimento di Fisica, Universit\`a di Lecce, Via Arnesano, CP 193, 73100
Lecce, and \\
INFN, Sezione di Lecce, Via Arnesano, CP 193, 73100 Lecce, Italy\\
$^3$
INFN, Sezione di Pavia, Via Bassi 6, I-27100, Pavia, Italy}

\date{ }
\maketitle 
 
\begin{abstract}
Microlensing events towards the Large Magellanic Cloud entail that 
a sizable fraction of dark matter is in the form of
MACHOs (Massive Astrophysical Compact Halo Objects), presumably located in the
halo of the Galaxy. Within the present uncertainties, brown
dwarfs are a viable candidate for MACHOs. Various reasons strongly
suggest that a large amount of MACHOs should actually consist of binary
brown dwarfs. Yet, this circumstance looks in flat contradiction with the
fact that MACHOs have been detected as  unresolved objects so far.
We show that such an apparent paradox does not exist within a model
in which MACHOs are clumped into dark clusters along with cold molecular
clouds, since dynamical friction on these clouds makes binary brown dwarfs
very close. 
Moreover, we argue that future microlensing experiments with a more 
accurate photometric observation can resolve binary brown dwarfs.
\end{abstract}

\section{Introduction}
As is well known (Trimble 1987; Carr 1994), 
the flat rotation curves in spiral galaxies supply the 
fundamental information that a large amount of dark matter is concealed in 
galactic halos, but unfortunately fail to provide any hint as regard to its 
constituents. 

As first proposed by Paczy\'nski (1986, 1996), gravitational 
microlensing can greatly help to clarify the nature of dark matter, 
and since 1993 this dream 
has started to become a reality with the detection of several microlensing 
events towards the Large Magellanic Cloud (Alcock et al. 1993; Aubourg
et al. 1993). 

Today, although the evidence for 
Massive Astrophysical Compact Halo Objects (MACHOs) -- presumably located in 
the halo of our galaxy -- is firm, the implications of such a discovery 
crucially depend on the assumed galactic model 
(Gates, Gyuk \& Turner 1996)\footnote{It has become customary  
to take the standard spherical 
halo model as a baseline for comparison. Unfortunately, because of 
the presently available limited statistics, different data-analysis procedures 
lead to results which are only marginally consistent.
Indeed, within the standard halo model, the average
MACHO mass value reported by the MACHO team is $0.5^{+0.3}_{-0.2}~M_{\odot}$ 
(Alcock et al. 1996), whereas the mass moment method (De R\'ujula, Jetzer \&
Mass\'o 1991) yields $0.27~M_{\odot}$ (Jetzer 1996).}. 
What can be reliably concluded from the existing data-set is that
MACHOs should lie in the mass range $0.05 - 1.0~M_{\odot}$,
but stronger claims are unwarranted because the MACHO mass strongly
depends on the uncertain properties of the considered galactic model (Evans
1996, De Paolis, Ingrosso \& Jetzer 1996).

High mass values $\ut > 0.5~M_{\odot}$ suggest white dwarfs as
a natural candidate for MACHOs (Tamanaha et al. 1990, Adams \&
Laughlin 1996, Chabrier, Segretein \& M\'era 1996, Fields, Mathews \&
Schramm 1996). Besides requiring a rather
{\it ad hoc} initial mass function sharply peaked somewhere
in the range $1-8~M_{\odot}$ and a halo age larger than $\sim 16$ Gyr,
this option implies an unobserved increased metallicity of the
interstellar medium (Gibson \& Mould 1996). 

Somewhat lower mass
values $\sim 0.3~M_{\odot}$ point towards the possibility that MACHOs are
hydrogen-burning main sequence M-dwarfs. 
However, the null results of several searches for
low-mass stars in our galaxy (Hu et al. 1994) entail that the
halo cannot be mostly in the form of M-dwarfs, and the same conclusion
is reached by optical imaging of high-latitude fields taken with the
Wide Field Camera of HST (Bahcall et al. 1994). Observe that
these results are derived under the assumption of a smooth spatial
distribution of M-dwarfs, and become considerably less severe 
in the case of a clumpy distribution (Kerins 1996, 1997). 

Still lower mass values
$\ut < 0.1~M_{\odot}$ make brown dwarfs an attractive candidate for MACHOs
\footnote{We stress that the limit for hydrogen burning -- usually quoted as
$0.08~M_{\odot}$ -- gets increased up to $0.11~M_{\odot}$ for
low-metallicity objects, such as a halo population (D'Antona 1987, 
Burrows, Hubbard \& Lunine 1989).}. Actually,
precisely the latter mass values are supported by nonstandard halo models
-- like the maximal disk model
(van Albada \& Sancisi 1986; Persic \& Salucci 1990) -- which tend to
be favoured by recent observational data (Sackett 1996).

In spite of the fact that present uncertainties 
do not permit to make any sharp statement
about the nature of MACHOs, brown dwarfs nevertheless look as a viable
possibility to date, and we shall stick to it throughout.

Once this idea is accepted, a few almost obvious consequences follow.  
Given the fact that ordinary halo stars form in (globular) clusters, it 
seems more likely that brown dwarfs as well form in clusters
(to be referred to as {\it dark clusters}) rather than in isolation. 
Furthermore --
since no known star-formation mechanism is very efficient -- we expect a
substantial fraction of the primordial gas to be left over. Finally --
because brown dwarfs do not give rise to stellar winds -- this gas should 
presumably remain confined within the dark clusters 
\footnote{
We emphasize that a halo primarily made of {\it unclustered} brown
dwarfs (as well as white and M-dwarfs) would contain too much 
diffuse hot gas (at virial temperature $\sim 10^6$ K) 
which emits in the X-ray band.}. As a matter of fact, it has been shown
(De Paolis et al. 1995a, 1995b; Gerhard \& Silk 1996) that these suggestions 
can be turned into a consistent scenario, which encompasses the 
Fall and Rees theory  for
the formation of globular clusters (Fall \& Rees 1985) 
and indeed predicts that dark clusters of brown dwarfs and 
cold molecular clouds -- mainly of $H_2$ -- should form in the halo at 
galactocentric distances larger than $10-20$ kpc.
A slightly different model based on the presence of a strong 
cooling-flow phase during the formation of our galaxy has been
considered by Fabian and Nulsen (1994, 1997) and leads to a halo made of
low-mass objects.

Before proceeding further, we would like to notice that the most
promising way to test whether MACHOs are indeed clumped into dark clusters
is through  
correlation effects in microlensing observations.
For the more massive dark clusters in the range $10^5 - 10^6 ~M_{\odot}$,
the implied 
degeneracy in the spatial and velocity distributions would result in a 
strong autocorrelation in the
sky position of microlensing events on an angular scale of $20\arcsec$,
along with a correlation in their duration.
It has been pointed out that
already a fairly poor statistics would be sufficient to rule out
the possibility that MACHOs are clumped into such massive clusters,
while to confirm it
more events are needed (Maoz 1994, Metcalf \& Silk 1996). On the other hand,
if either the mass of the dark clusters is considerably smaller than 
$10^6 M_{\odot}$ or the fraction of dark matter in the 
form of MACHOs is small, then a much larger
sample of events may be required.

The aim of the present paper is to address a further issue that the above 
picture implies. As we shall discuss below, a large fraction of MACHOs
is expected to consist of binary brown dwarfs. However, this circumstance is 
an occasion for anxiety, since the overwhelming majority of MACHOs looks
like unresolved objects in current microlensing experiments.
We will demonstrate that such an apparent paradox does not actually exist 
within the model in question. For, dynamical friction on molecular clouds
makes binary brown dwarfs so close that they cannot have been resolved so far
-- still, we argue that they can be resolved in future 
microlensing experiments with a more accurate photometric observation.

\section{Preliminary considerations}
A thorough account of the scenario under consideration has been reported
elsewhere, along with an analysis of several observational tests 
(De Paolis et al. 1996).
We should stress that the lack of observational informations about
dark clusters would make any effort to understand their structure and dynamics 
hopeless, were it not for some remarkable insights that our 
unified treatment of globular and dark clusters provides us. 

In the first place,
it looks quite natural to assume that also dark clusters have a denser core
surrounded by an extended spherical halo. Moreover, 
it seems reasonable to suppose (at least tentatively)
that dark clusters have the same average mass density as globular clusters.
Accordingly, we get $r_{DC} \simeq 0.12~(M_{DC}/M_{\odot})^{1/3}$ pc,
with $M_{DC}$ and $r_{DC}$ denoting the mass and median 
radius of a dark cluster, respectively. So, by the virial theorem the
one-dimensional velocity dispersion $\sigma$ of MACHOs and molecular
clouds within a dark cluster reads
\begin{equation}
\sigma\simeq 6.9\times
10^{-2}~\left(\frac{M_{DC}}{M_{\odot}}\right)^{1/3}
~{\rm km~s^{-1}}~.
\label{sigma*}
\end{equation}
In addition, dark clusters -- just like globular clusters -- presumably
stay for a long time in a quasi-stationary phase, with
an average central density $\rho(0)$ slightly lower than 
$10^4~M_{\odot}$ pc$^{-3}$ (which is the observed 
average central density for globular clusters).

Below, we recall only some basic points of the considered picture, which
are important for the considerations to follow.
First, the amount of virialized diffuse gas inside dark clusters should be
small -- otherwise it should 
have been observed in the radio band --
and so most of the gas has to be in the form of 
cold self-gravitating clouds clumped into dark clusters. Second -- at
variance with the case of globular clusters -- the mass distribution of
dark clusters should be smooth, with a corresponding spectrum
extending from $M_{DC} \sim 10^6 ~M_{\odot}$ down to much lower values. 
Third, the dark 
clusters with  $3 \times 10^2 ~M_{\odot} \ut < M_{DC} \ut < 10^6 ~M_{\odot}$
are expected to have survived all disruptive effects and should
still populate the outer galactic halo today. Fourth, the dark clusters with
$M_{DC} \ut < 5 \times 10^4 ~M_{\odot}$ should have begun
core collapse.
Fifth --
since the temperature of these clouds is close to that of the Cosmic 
Background Radiation (De Paolis et al. 1995c) -- 
the virial theorem entails
\begin{equation}
r_m\simeq 4.8\times 10^{-2}\left( \frac{M_m}{M_{\odot}}\right)~{\rm pc}~,
\label{rm}
\end{equation}
where $r_m$ and $M_m$ indicate the median radius and mass, respectively, 
of a self-gravitating cloud.
Accordingly, the average number density $n_m$ in a cloud turns out to
be 
\begin{equation}
n_m \simeq 62.2 \left(\frac{\rm pc}{r_m} \right)^2~ {\rm cm^{-3}}~.
\label{3}
\end{equation}
Sixth, the leftover gas is mainly $H_2$.

We suppose for definiteness (and with an eye to microlensing
experiments) that all individual brown dwarfs have mass 
$m\simeq 0.1~M_{\odot}$.
Furthermore, since also molecular clouds originate from the fragmentation
process which produces the brown dwarfs, we suppose (for definiteness)
that they lie in the mass range
$10^{-3}~M_{\odot}\ut < M_m \ut < 10^{-1}~M_{\odot}$. 
Correspondingly, eqs. (\ref{rm}) and (\ref{3}) yield
$4.8 \times 10^{-5}~ {\rm pc} \ut < r_m \ut <
4.8 \times 10^{-3}~{\rm  pc}$ and $2.7 \times 10^{10}~{\rm cm^{-3}} 
\ut > n_m  \ut > 2.7 \times 10^6~{\rm cm^{-3}}$, respectively.

\section{MACHOs as binary brown dwarfs}
Much in the same way as it occurs for ordinary stars, also in the
present case the fragmentation mechanism that gives rise to individual
brown dwarfs should presumably produce a substantial fraction of binary
brown dwarfs  -- they will be referred to as {\it primordial} binaries.
It is important to keep in mind that their mass fraction 
$f_{PB}$ can be as large as 50\% and that -- 
because of the mass stratification instability -- 
they will concentrate inside the dark cluster 
cores (Spitzer 1987), which are therefore expected to be chiefly composed
by binaries and molecular clouds.
In addition -- as far as dark clusters with $M_{DC} \ut < 5 \times
10^4 M_{\odot}$ are concerned -- {\it tidally-captured} binary brown dwarfs
ought to form in the cores because of the increased central 
density caused by core collapse
(Fabian, Pringle \& Rees 1975). Thus, we are led to the conclusion 
that MACHOs should consist of both {\it individual} and {\it binary} 
brown dwarfs within the present picture.

Our subsequent analysis will be carried out in the general case of
arbitrary $f_{PB}$, but our interest is of course focused on the
case of large $f_{PB}$.

We recall that a binary system is hard when its 
binding energy exceeds the kinetic energy of field stars (otherwise it is
soft). 
As is well known, soft binaries always get softer whereas hard binaries always
get harder because of encounters with individual stars (Heggie 1975). 
In the case under consideration, binary brown dwarfs happen
to be hard when their orbital radius $a$ obeys the constraint
\begin{equation}
a \ut < 1.4 \times 10^{12} \left( \frac{M_{\odot}}{M_{DC}} \right)^{2/3}
{\rm km}~.
\label{eqno:1}
\end{equation}
Hence, only those binaries which satisfy condition (\ref{eqno:1}) 
can survive up until the present.

Now,  consistency with the results of microlensing
experiments -- in the sense specified above -- requires that the present
orbital radius of the overwhelming majority of binary brown dwarfs
should be (roughly) less than one-half of
the Einstein radius for microlensing towards the LMC 
(Gaudi \& Gould 1996).
This demand  implies in turn
that the further constraint 
\begin{equation}
a_{today} \ut < 3 \times 10^8~{\rm km}
\label{5} 
\end{equation}
has to be met today.
What is crucial to realize 
is that the latter bound turns out to be {\it stronger}
than the former for $M_{DC}\ut <3.2\times 10^5~M_{\odot}$.

Let us consider first {\it tidally-captured} binaries, 
whose formation is ultimately
brought about by core collapse in dark clusters and is expected to 
proceed as in the case of globular clusters. 
So, following the analysis of Press and Teukolsky
(1977) and Lee and Ostriker (1986) we deduce 
that practically all individual brown dwarfs in a dark cluster
core get captured
into binaries as soon as the
core density \footnote{We suppose throughout that the core of dark clusters 
has a constant density profile.} 
$\rho(0)$ starts to satisfy the condition
\begin{equation}
\rho(0) \ut > \frac{3.2 \times 10^4}{f_{IBD}} \left( \frac{M_{DC}}{M_{\odot}} 
\right)^{0.4} M_{\odot} ~{\rm pc}^{-3}~,
\label{eqno:2}
\end{equation}
where $f_{IBD}$  denotes the mass fraction of individual brown dwarfs
in the core. 
According to 
the above assumptions, we expect $\rho(0) \simeq 10^4 ~M_{\odot}$ pc$^{-3}$
{\it just before core collapse}, and so 
a moderate increase in $\rho(0)$ may be sufficient to make 
tidal capture operative -- 
hence this process would occur at the onset of core collapse. 
However, this conclusion depends on $f_{PB}$, since $f_{IBD}$
necessarily gets small for large $f_{PB}$.
What is
the orbital radius of tidally-captured binary brown dwarfs? Following the
procedure outlined by Statler, Ostriker and Cohn (1987) we easily find 
$a \simeq 2.5 \times 10^5$ km (this value is almost independent of
$M_{DC}$), and so tidally-captured binaries are very hard and obviously
obey condition (\ref{5}). Unfortunately, tidally-captured binary brown 
dwarfs are irrelevant from the observational point of view, since their 
fractional abundance turns out to be anyway less than 1\%.

What is the fate of
{\it primordial} binaries? Surely, only those which are
hard can survive. In fact, were individual - binary encounters the
only
relevant process, we would conclude that hard binary brown dwarfs should
indeed survive. However, also binary - binary encounters play an
important r\^ole in the dark cluster cores, where binaries are far more
abundant than individual brown dwarfs as long as
$f_{PB}$ is not negligibly small.
Now, it is well known that in the latter process one of two binaries
gets often disrupted (this cannot happen for both binaries -- given
that they are hard -- while fly bys are rather infrequent), thereby leading 
to the depletion of the primordial binary population.
We will discuss this effect in Sect. 5, where we shall find that for
realistic values of the dark cluster parameters the binary break up does
not take place.
Still, this is not the end of the story. For, all values for their
orbital radius consistent with condition (\ref{eqno:1}) are in
principle allowed. Thus, it is crucial that primordial binaries manage
to shrink in such a way that {\it also condition} (\ref{5}) 
{\it is eventually met}.

\section{Hardening processes}
Let us turn our attention to the specific mechanism by which
condition (\ref{5}) can get fulfilled. Observe that this requirement 
demands that a binary should give up a binding energy
larger than $4.5 \times 10^{43}$ erg.

\subsection{Collisional hardening} 
Collisional hardening -- namely the process whereby hard
binaries get harder in encounters with individual brown dwarfs --
looks as the most natural possibility. In order to see whether it
works, we consider the associated average hardening rate (Spitzer
\& Mathieu 1980), which reads presently
\begin{equation}
< \dot E>_{CH}~ \simeq~ -~2.8~ \frac{G^2 m^3~ n_{IBD}(0)}{\sigma}~,
\label{11}
\end{equation}
where $n_{IBD}(0) \simeq f_{IBD}~ \rho(0)/m$ is the number density of
individual brown dwarfs in the core. So, eq. (\ref{11}) becomes
\begin{eqnarray}
<\dot E>_{CH}~  &\simeq & ~ -~1.7 \times 10^{32} f_{IBD}
\left(\frac{M_{\odot}}{M_{DC}} \right)^{1/3}\times \nonumber \\ 
& & \left(\frac{\rho(0)}{M_{\odot} ~{\rm pc}^{-3}} \right) 
~~{\rm erg}~{\rm yr}^{-1}~,
\label{edot}
\end{eqnarray}
thanks to eq. (\ref{sigma*}). 
From eq. (\ref{edot}) we can estimate the total amount of binding
energy $\Delta E_B$ released by a primordial binary via encounters with
individual brown dwarfs during the Universe lifetime \footnote{The age
of the Universe is nominally taken to be $10^{10}$ yr throughout.}.
Taking $\rho (0) \sim 10^3~M_{\odot}$ pc$^{-3}$ -- in agreement with
our assumptions -- we find 
$\Delta E_B \sim 10^{43} - 10^{44}~ f_{IBD}$ erg (depending on $M_{DC}$).
As already pointed out, a large $f_{PB}$ would imply a small $f_{IBD}$,
in which case 
collisional hardening fails to reduce the
present orbital radius of primordial binary brown dwarfs down to a
value consistent with condition (\ref{5}). Furthermore, denoting 
by $\Delta a$ the change of orbital radius corresponding to a loss
of binding energy $\Delta E_B$, we have $\mid \Delta E_B/E_B \mid=
\mid \Delta a /a \mid$, which shows more generally that collisional
hardening turns out to be
irrelevant for microlensing observations.

\subsection{Frictional hardening}
As we shall show below, the presence of molecular 
clouds in the dark cluster cores
-- which  is indeed the most characteristic feature of the model in
question -- provides
a novel mechanism by which primordial binary brown dwarfs give up a
larger amount of binding energy, thereby getting so hard that
condition (\ref{5}) is actually obeyed.
Basically, this occurs because of
dynamical friction  on molecular clouds, and so it will be referred to
as {\it frictional hardening}. 

It is not difficult to extend the standard treatment of dynamical friction
(Binney \& Tremaine 1987) to the relative motion of the
brown dwarfs in a binary system which moves inside a molecular cloud.
For simplicity, we assume that molecular clouds have a constant density
profile $\rho_m$. In the case of a circular orbit \footnote{Indeed, the
circularization of the orbit is achieved by tidal effects after a
few periastron passages (Zahn 1987).}, 
the equations of motion imply that the time $t_{21}$ needed to reduce the
orbital radius $a$ from $a_1$ down to $a_2$ is
\begin{equation}
t_{21} \simeq 0.17\left(\frac{m}{G}\right)^{1/2}
\frac{1}{\rho_m\ln\Lambda}~(a_{2}^{-3/2}-a_{1}^{-3/2})~,
\label{t21}
\end{equation}
where the Coulomb logarithm reads
$\ln\Lambda\simeq \ln ({r_m v_c^2}/{Gm})\simeq\ln ({r_m}/{a_1})$,
with $v_c$ denoting the circular velocity (approximately
given by Kepler's third law). Manifestly, the diffusion approximation -- 
upon which the present treatment is based -- requires that the orbital 
radius of a binary should always be smaller than the median radius of a 
cloud.
As we are concerned henceforth with hard
binaries, $a_1$ has to obey condition (4). On the other hand, $a_1$ is
the larger value for the orbital radius in eq. (\ref{t21}). So, we
shall take for definiteness -- in the Coulomb logarithm only -- 
$a_1\simeq 1.4\times 10^{12}(M_{\odot}/M_{DC})^{2/3}$ km. In addition,
from eq. (2) we have 
$\rho_m\simeq 2.5 ~({\rm pc}/r_m)^2~M_{\odot}~{\rm pc}^{-3}$.
Hence, putting everything together we obtain
\begin{equation}
t_{21} \simeq 5\times 10^{25}~\Xi\left(\frac{r_m}{{\rm pc}}\right)^2
\left[\left(\frac{{\rm km}}{a_2}\right)^{3/2} -
\left(\frac{{\rm km}}{a_1}\right)^{3/2}
\right]~{\rm yr},
\label{t21new}
\end{equation}
having set
\begin{equation}
\Xi\equiv [3+\ln({r_m}/{{\rm pc}})+0.7\ln({M_{DC}}/{M_{\odot}})]^{-1}~.
\label{15}
\end{equation}
Specifically, the diffusion approximation demands $\Xi >0$, which yields in 
turn
\begin{equation}
r_m >5\times 10^{-2}\left(\frac{M_{\odot}}{M_{DC}}\right)^{0.7}~~{\rm pc}~.
\end{equation}
Observe that for $M_{DC}\ut < 2.1\times 10^4~M_{\odot}$ this constraint
restricts the range of allowed values of $r_m$ as stated in Sect. 2.

Let us begin by estimating the time $t_{\rm frict}$ 
required by dynamical friction to 
reduce the orbital radius from 
$a_{\rm in}\simeq 1.4\times 10^{12}~(M_{\odot}/M_{DC})^{2/3}$ km down to
$a_{\rm fin}\simeq 3\times 10^{8}$ km, under the assumption
that a binary moves all the time inside molecular 
clouds. Clearly, from eq. (\ref{t21new}) we get
\begin{equation}
t_{\rm frict} \simeq 9.5\times 10^{12}~\Xi\left(\frac{r_m}{{\rm
pc}}\right)^2 A~~{\rm yr}~,
\label{tfrict}
\end{equation}
where we have set
\begin{equation}
A \equiv 
\left[1-3.1\times 10^{-6}\left(\frac{M_{DC}}{M_{\odot}}\right)\right]~.
\end{equation}
As an indication, we notice that for $r_m\simeq 10^{-3}$ pc 
($M_m \simeq 2 \times 10^{-2} ~M_{\odot}$) and
$M_{DC}\simeq 10^5~M_{\odot}$ we find $\Xi\simeq 0.2$ and so 
$t_{\rm frict}\simeq 1.3\times 10^6$ yr. 

Were the dark clusters completely filled by clouds, eq. (\ref{tfrict}) 
would be the 
final result. However, the distribution of the clouds is lumpy. So, 
$t_{\rm frict}$ has to be understood as the total time spent by a binary 
inside the clouds. Thus, the total time demanded to reduce $a_{in}$ down to 
$a_{fin}$ is evidently longer than $t_{\rm frict}$ and can be computed by 
the following procedure. Keeping in mind that both the 
clouds and the binaries have
average velocity $v\simeq \sqrt{3}\sigma$ (for simplicity, we
neglect the equipartition of kinetic energy of the binaries)
it follows that the time needed by a binary to cross a single cloud is
\begin{equation}
t_m\simeq \frac{r_m}{\sqrt{2}~v}\simeq
5.6\times 10^6\left(\frac{r_m}{{\rm pc}}\right)
\left(\frac{M_{\odot}}{M_{DC}}\right)^{1/3}~{\rm yr}~.
\label{16}
\end{equation}
We see that the above values of $r_m$ and $M_{DC}$ entail
$t_m\simeq 1.2\times 10^2$ yr. Therefore, frictional  
hardening involves many clouds and is actually 
accomplished only after
\begin{equation}
N_m\simeq\displaystyle{\frac{t_{\rm frict}}{t_m}}
\simeq 1.7\times 10^6~\Xi
\left(\displaystyle{\frac{r_m}{{\rm pc}}}\right)
\left(\frac{M_{DC}}{M_{\odot}}\right)^{1/3} A
\label{tildeN1}
\end{equation}
clouds have been traversed. Taking again  the above values for $r_m$ and
$M_{DC}$, we get $N_m\simeq 1.1\times 10^4$.

It is illuminating to find how many crossings of the core are necessary
for a binary to traverse $N_m$ clouds.
To this end, we proceed to estimate the number of clouds $N_c$ 
encountered during {\it one} crossing of the core. 
Describing the dark clusters by a King model, we can identify the core
radius with the King radius. Evidently, the cross-section
for binary - cloud encounters is $\pi r_m^2$ and so we have
\begin{equation}
N_c\simeq
\left(\frac{9\sigma^2}{4\pi G\rho(0)}\right)^{1/2} n_{CL}(0)~ \pi~r_m^2~,
\label{tildeN2}
\end{equation}
with $n_{CL}(0)$ 
denoting the cloud number density in the core.
Thanks to eq. (\ref{rm}), we can write
\begin{eqnarray}
n_{CL}(0) & \simeq & f_{CL}\frac{\rho(0)}{M_m} \nonumber \\
&\simeq &
4.8\times 10^{-2}f_{CL}\left(\frac{{\rm pc}}{r_m}\right)\times \nonumber \\
& &
\left(\frac{\rho(0)}{M_{\odot}~{\rm pc}^{-3}}\right)~{\rm pc}^{-3}~,
\end{eqnarray}
where $f_{CL}$ denotes the fraction of core dark matter in the form of 
molecular clouds. Correspondingly, eq. (\ref{tildeN2}) becomes
\begin{eqnarray}
N_c &\simeq & 0.13 \times f_{CL}
\left(\displaystyle{\frac{\rho(0)}{M_{\odot}~{\rm
pc}^{-3}}}\right)^{1/2}
~\times \nonumber\\
& &
\left(\displaystyle{\frac{M_{DC}}{M_{\odot}}}\right)^{1/3}
\left(\displaystyle{\frac{r_m}{\rm pc}}\right)~,
\end{eqnarray}
on account of eq. (1).
So, the total number of core crossings $N_{cc}$ that a binary 
has to make in order to reduce its orbital radius down to
$a_{\rm fin} \simeq 3 \times 10^8$ km is 
\begin{equation}
N_{cc}\simeq\displaystyle{
\frac{N_m}{N_c}} \simeq 1.3\times 10^7~ f_{CL}^{-1}~
\Xi 
\left(\displaystyle{\frac{M_{\odot}~{\rm
pc}^{-3}}{\rho(0)}}\right)^{1/2} A~.
\end{equation}
On the other hand, the core crossing time is 
\begin{eqnarray}
t_{cc} & \simeq &
\left(\frac{9\sigma^2}{4\pi G\rho(0)}\right)^{1/2}\frac{1}{v}
\nonumber \\
& \simeq & 7\times 10^6\left(\frac{M_{\odot}~{\rm pc}^{-3}}{\rho(0)}
\right)^{1/2}~{\rm yr}~.
\end{eqnarray}
Thus, the total time needed by frictional hardening to make primordial
binary brown dwarfs unresolvable in present microlensing experiments is
\begin{eqnarray}
t_{\rm tot}  &\simeq & 
N_{cc}~t_{cc} 
\simeq  9 \times 10^{13}
f^{-1}_{CL}\times \nonumber \\
& & ~\Xi
\left(\frac{M_{\odot}~{\rm pc}^{-3}}{\rho(0)} \right ) ~A
~~~{\rm yr} ~.
\label{58}
\end{eqnarray}
Observe that $t_{\rm tot}$ is insensitive to $r_m$ and
almost insensitive to $M_{DC}$ (this
dependence enters only logarithmically in $\Xi$).
As a matter of fact, since we have considered so far only the 
extreme case of largest
$a_{\rm in}$, $t_{\rm tot}$ in eq. (\ref{58}) should be understood as an 
{\it upper bound} to the total hardening time. 
Accordingly --
taking $f_{CL} \simeq 0.5$ -- we see that for the above
illustrative values of $r_m$ and $M_{DC}$ we get 
$t_{\rm tot}$ shorter than the age of the Universe
for $\rho(0) \ut > 2.5 \times 10^3~M_{\odot}~ {\rm pc}^{-3}$.
As the latter value is consistent with our assumptions, we conclude that 
frictional hardening has enough time to operate.

\section{Binary - binary encounters}
We are now in position to investigate the implications of 
binary - binary encounters.

As a first step, we recall that the average
rate $\Gamma$ for individual - binary (IB) and binary - binary (BB) encounters
(Spitzer \& Mathieu 1980) can presently be written as
\begin{equation}
\Gamma ~ \simeq~ \alpha~ \frac{G m a}{\sigma}~,
\label{7}
\end{equation}
with $\alpha$ being either $\alpha_{IB} \simeq 14.3$ 
or $\alpha_{BB} \simeq 13$. Correspondingly, on account of eq. (\ref{sigma*})
the reaction time in the dark cluster cores turns out to be
\begin{equation}
t_{\rm react} \simeq 10^{19} \beta \left(\frac{M_{\odot}~{\rm pc}^{-3}}
{\rho(0)} \right) \left(\frac{M_{DC}}{M_{\odot}} \right)^{1/3} 
\left( \frac{{\rm km}}{a} \right) ~{\rm yr}~,
\label{60}
\end{equation}
with $\beta$ being either $\beta_{IB} \simeq 3.1/f_{IBD}$ or 
$\beta_{BB} \simeq 6.7/f_{PB}$. 
Since both processes are {\it a priori} of comparable strength and 
in the cores we expect $f_{IBD} \ll f_{PB}$, individual - binary encounters
will be neglected.

As already stressed, binary - binary encounters can lead to their disruption.
Now, if no hardening were to occur -- which means that the orbital 
radius $a$ would stay constant -- the binary survival condition would
simply  follow by demanding that $t^{BB}_{react}$ should exceed the age
of the Universe.
However, hardening makes $a$ decrease, and so $t^{BB}_{react}$ increases
with time. This effect can be taken into account by considering the
averge value $< t^{BB}_{react} >$ of the reaction time over the time
interval in question (to be denoted by T), namely 
\begin{equation}
< t^{BB}_{react} >~ \equiv ~\frac{1}{T} ~ \int_0^T dt~t^{BB}_{react}~.
\label{aver}
\end{equation}
In order to compute $< t^{BB}_{react} >$ the temporal dependence
$a(t)$ of the binary orbital radius is required. This quantity can
be obtained as follows. We have seen that the total time $t^{TOT}_{21}$
needed to reduce $a_1$ down to $a_2$ is actually longer than $t_{21}$,
since the cloud distribution is lumpy. So, by going through the same 
steps as before, we easily find
\begin{eqnarray}
t^{TOT}_{21} & \simeq & 4.8\times 10^{26}~\Xi~ f^{-1}_{CL}
\left(\frac{M_{\odot} {\rm pc^{-3}}}
{\rho(0)}\right) \times \nonumber \\ & &
\left[\left(\frac{{\rm km}}{a_2}\right)^{3/2} -
\left(\frac{{\rm km}}{a_1}\right)^{3/2}
\right]~{\rm yr}~.
\label{t21tot}
\end{eqnarray}
Setting for notational convenience $t \equiv t^{TOT}_{21}$, 
$a_0 \equiv a_1$ and $a(t) \equiv a_2$, eq. (\ref{t21tot}) yields 
\begin{eqnarray} 
a(t) & \simeq & \left[ \left(\frac{{\rm km}}{a_0}\right)^{3/2}
+ 2.1 \times 10^{-27}~ \Xi^{-1} f_{CL} \right.
\times \nonumber \\
& & \left. \left(\frac{\rho(0)}{M_{\odot}
{\rm pc^{-3}}} \right)\left(\frac{t}{\rm yr}\right ) \right]^{-2/3}
~{\rm km}~.
\label{n1}
\end{eqnarray}
Combining eqs. (\ref{60}) and (\ref{n1}) together and inserting the
ensuing expression into eq. (\ref{aver}), we get
\begin{eqnarray}
< t^{BB}_{react} > \simeq 1.9 \times 10^{46}~ f^{-1}_{PB}
f^{-1}_{CL}~ \Xi \left( \frac{M_{DC}}{M_{\odot}}\right)^{1/3} 
\left(\frac{{\rm yr}}{T}\right) 
\times \nonumber \\
\left(\frac{\rm km}{a_0} \right)^{5/2}
\left(\frac{M_{\odot} {\rm pc^{-3}}}{\rho(0)}\right)^2
\left\lbrace \left[1+2.1\times 10^{-27} \Xi^{-1} f_{CL} \right. \right.
\times \nonumber \\ 
\left. \left.
\left(\frac{\rho(0)}{M_{\odot} {\rm pc^{-3}}}\right)
\left(\frac{T}{\rm yr}\right) \left(\frac{a_0}{\rm km}\right)^{3/2}
\right]^{5/3} -1 \right\rbrace~{\rm yr}~.
\label{n2}
\end{eqnarray} 
Let us now require $< t^{BB}_{react} >$ to exceed the age of the
Universe (taking evidently $T \simeq 10^{10}$ yr). As it is apparent
from eq. (\ref{60}), $t^{BB}_{react}$ is shorter for softer binaries. 
Hence, in order to contemplate {\it hard} binaries with an arbitrary orbital
radius we have to set $a_0 \simeq 1.4 \times 10^{12} (M_{\odot}/M_{DC})^{2/3}$
km in eq. (\ref{n2}). Correspondingly, the binary survival condition
reads \footnote{Application of the same argument to tidally - captured
binaries shows that no depletion occurs in this way.}
\begin{eqnarray}
f_{PB}  \ut <  8.2 \times 10^{-5} f_{CL}^{-1}~ \Xi
\left(\frac{M_{DC}}{M_{\odot}}\right)^2 \left( \frac{M_{\odot}{\rm pc^{-3}}}
{\rho(0)} \right)^2 \times \nonumber \\
\left\lbrace
\left[1+ 35 f_{CL} \Xi^{-1} \left(\frac{\rho(0)}{M_{\odot}{\rm pc^{-3}}}
\right) \left(\frac{M_{\odot}}{M_{DC}}\right)\right]^{5/3} -1
\right\rbrace~.
\label{n3}
\end{eqnarray}
Although the presence of various dark cluster parameters prevents a 
clear - cut conclusion to be drawn from eq. (\ref{n3}), in the
illustrative case $M_{DC} \simeq 10^5 M_{\odot}$ and $f_{CL} \simeq 0.5$
eq. (\ref{n3}) entails e.g. $f_{PB} \ut < 0.3$ for
$\rho(0) \simeq 3 \times 10^3 M_{\odot}~ {\rm pc^{-3}}$. Thus, we infer
that for realistic values of the parameters in question a sizable fraction of 
primordial binary brown dwarfs survives binary - binary encounters in the
core \footnote{If the initial value of $f_{PB}$ fails to satisfy 
condition (\ref{n3}), primordial binaries start to be destroyed in binary - 
binary encounters until their fractional abundance gets reduced down to a value
consistent with condition (\ref{n3}).}.

Binary - binary encounters also have an implication of direct relevance
for frictional hardening. Indeed, before claiming that such a mechanism
really 
accomplishes its job, we have to make sure that primordial
binaries do not leave the core before their orbital radius is 
reduced down to acceptable values. As in the case of globular clusters,
encounters between individual and binary (IB) brown dwarfs can
give them enough 
kinetic energy to escape from the core, and the same happens in
encounters between binaries (BB) (Spitzer 1987).
As explained above, we can restrict our attention to the latter 
process, and we consider $< t^{BB}_{react} >$ as averaged over
the total hardening time $t_{tot}$ (so we are going to take
$T \simeq t_{tot}$ in eq. (\ref{n2})). Again, in order to contemplate 
{\it hard} binaries with an arbitrary orbital radius we set 
$a_0 \simeq 1.4 \times 10^{12} (M_{\odot}/M_{DC})^{2/3}$ km 
in eq. (\ref{n2}). Recalling eq. (\ref{58}), we find
\begin{eqnarray}
\frac{< t^{BB}_{react} >}{t_{tot}} \simeq  10^{-12} f_{CL} f_{PB}^{-1}
\Xi^{-1} A^{-2} \left(\frac{M_{DC}}{M_{\odot}}\right) \times \nonumber \\
\left\lbrace\left[1 + 3.1 \times 10^5 A \left(\frac{M_{\odot}}{M_{DC}}\right)
\right]^{5/3} -1 \right\rbrace~{\rm yr}~.
\label{n4}
\end{eqnarray}
As in the case of eq. (\ref{n3}), the various parameters showing up in 
eq. (\ref{n4}) do not permit to make sharp statements, but in the
illustrative case $M_{DC} \simeq 10^5 ~M_{\odot}$, $f_{CL} \simeq 0.5$
and $f_{PB} \ut < 0.3$ eq. (\ref{n4}) yields $< t^{BB}_{react} >
\ut > t_{tot}$. In conclusion, it seems fair to say that condition
(5) is ultimately met by the overwhelming majority of binary brown
dwarfs (this statement is especially true in view of the fact that 
$t_{tot}$ --
as given by eq. (\ref{58}) -- has to be viewed as an upper bound
to the total hardening time).

\section{Orbital radius of primordial binaries}
Our concern was primarily addressed to show that primordial binaries are
hard enough today, so as to avoid conflict with the results of 
microlensing experiments. Still, it would be important to know how large
is their present orbital radius. Unfortunately, the lack of information
about the  formation mechanism prevents a clear-cut answer to be given,
but it looks remarkable that a nontrivial -- and in fact very 
interesting -- lower bound can nevertheless be derived.
Basically, the argument is very similar to the previous one
that led us to eq. (\ref{t21tot}). However, it will now be used
with a different logic, for we shall leave both $a_1$ and $a_2$ free while
demanding that 
$t_{21}^{TOT}$ should equal the age of the Universe. Correspondingly,
eq. (\ref{t21tot}) implies
\begin{eqnarray}
\left(\frac{{\rm km}}{a_2}\right)^{3/2} & \simeq &  
\left(\frac{{\rm km}}{a_1}\right)^{3/2} +2.1\times 10^{-17}f_{CL}
\times \nonumber \\
& & \Xi^{-1}
\left(\frac{\rho(0)}{M_{\odot}~{\rm pc}^{-3}}\right )~.
\label{31}
\end{eqnarray}
Manifestly, frictional hardening is operative to the extent that $a_2$
becomes considerably smaller than $a_1$. Accordingly, from eq. (\ref{31}) we
see that this is indeed the case provided
\begin{equation}
a_1  \ut  > 1.3\times 10^{11} f_{CL}^{-2/3}~\Xi^{2/3}
\left(\frac{M_{\odot}~{\rm pc}^{-3}}{\rho(0)}\right)^{2/3}
~~{\rm km}~.
\label{100}
\end{equation}
Owing to eq. (\ref{100}), eq. (\ref{31}) entails
\begin{equation}
a_2 \simeq  1.3\times 10^{11} f_{CL}^{-2/3}~\Xi^{2/3}
\left(\frac{M_{\odot}~{\rm pc}^{-3}}{\rho(0)}\right)^{2/3}~~{\rm km}~.
\label{33}
\end{equation}
Physically, the emerging picture is as follows. Only those primordial
binaries whose initial orbital radius satisfies condition (\ref{100}) are
affected by frictional hardening, and their present orbital radius
turns out to be almost {\it independent} of the initial
value. We can make the present discussion somewhat more specific by
noticing that our assumptions strongly suggest
$\rho(0)\ut < 10^4~M_{\odot}~{\rm pc}^{-3}$, in which case
both eq. (\ref{100}) and eq. (\ref{33}) acquire the form
\begin{equation}
a_{1,2}\ut > 2.8 \times 10^{8} f_{CL}^{-2/3}\Xi^{2/3}~{\rm km}~.
\label{34}
\end{equation}
Evidently, very hard primordial
binaries -- which violate condition (\ref{34}) -- do not suffer frictional
hardening, and the same is true for tidally-captured binaries. We
stress that these conclusions are (practically) unaffected by
collisional hardening.

\section{Energy Balance}
As the above analysis shows, dynamical friction transfers a huge amount of 
energy from primordial binary brown dwarfs to molecular clouds, and so
it looks compelling to investigate (at least) the gross features of
the energy balance.

Let us start by evaluating the energy  acquired by molecular
clouds in the process of frictional hardening. Recalling that the traversal
time for a single cloud is given by eq. (15), eq. (\ref{t21new}) 
entails that --
after a binary with initial orbital radius $a_1$ 
has crossed $N$ clouds -- its orbital radius gets reduced down to
\begin{equation}
{a_{N+1}} \simeq \left[ 
B_N
\left(\frac{M_{\odot}}{M_{DC}}\right)^{1/3} + 
\left(\frac{{\rm km}}{a_1}\right)^{3/2} \right]^{-2/3}~{\rm km}~, 
\label{29}
\end{equation}
with $B_N\equiv 1.1\times 10^{-19}N~\Xi^{-1}({\rm pc}/{r_m})$.
Accordingly, we see that the orbital radius remains almost constant
until $N$ reaches the critical value
\begin{equation}
N_*\equiv 9\times 10^{18}~\Xi
\left(\frac{r_m}{{\rm pc}} \right)
\left(\frac{M_{DC}}{M_{\odot}} \right)^{1/3}
\left(\frac{{\rm km}}{a_1}\right)^{3/2}~,
\end{equation}
whereas it {\it decreases} 
afterwards. Because the
energy acquired by the clouds is just the binding energy given up by 
primordial binaries, the above information can be directly used to
compute the energy $\Delta E_c(N)$ gained by the $N$-th cloud traversed
by a binary with initial orbital radius $a_1$. Manifestly, we have
\begin{equation}
\Delta E_c(N)=\frac{1}{2}Gm^2
\left(\frac{1}{a_{N+1}}-\frac{1}{a_N} \right)~.
\end{equation}
Thanks to eq. (\ref{29}), 
a straightforward calculation shows that $\Delta E_c(N)$
stays practically constant
\begin{eqnarray}
\Delta E_c & \simeq & 9.8\times 10^{32}~\Xi^{-1}
\left(\frac{{\rm pc}}{r_m} \right)\times \nonumber \\
& & \left(\frac{M_{\odot}}{M_{DC}} \right)^{1/3}
\left(\frac{a_1}{{\rm km}}\right)^{1/2}~~{\rm erg}~,
\label{32}
\end{eqnarray}
as long as $N\ut < N_*$, while it subsequently {\it decreases}. 
So, the
amount of energy transferred to a cloud is maximal during the early stages 
of hardening. Now, since the binding energy of a cloud is
\begin{equation}
E_c\simeq 7.7\times 10^{42}
\left(\frac{r_m}{{\rm pc}} \right)~~{\rm erg}~,
\end{equation}
it can well happen that $\Delta E_c >E_c$ (depending on $r_m$, $M_{DC}$ 
and $a_1$), which means that the cloud would evaporate unless it
manages to efficiently dispose of the excess energy.

A deeper insight into this issue can be gained as follows
(we focus the attention on the early stages of hardening, when the
effect under consideration is more dramatic).
Imagine that a spherical cloud is crossed by a primordial 
binary which moves along a straight line, and consider the cylinder $\Delta$
traced by the binary inside the cloud (its volume being approximately
$\pi a^2 r_m$). Hence, by eq. (3) the average
number of molecules inside $\Delta$ turns out to be
\begin{equation}
N_{\Delta} \simeq 5.8 \times 10^{30} 
\left( \frac{a_1}{{\rm km }}\right)^2
\left( \frac{{\rm pc}}{ r_m}\right)~.
\label{NDelta}
\end{equation}
Physically, the energy $\Delta E_c$ gets first deposited within $\Delta$
in the form of heat. Neglecting thermal conductivity (more about this,
later), the temperature inside $\Delta$ accordingly becomes
\begin{eqnarray}
T_{\Delta} & \simeq & \frac{2}{3}\frac{\Delta E_c}{N_{\Delta}k_B} \\ \nonumber
& \simeq & 8.1 \times 10^{17}~\Xi^{-1} 
\left( \frac{M_{\odot}}{M_{DC}}\right)^{1/3}
\left( \frac{{\rm km }}{a_1}\right)^{3/2}~~{\rm K}~,
\label{Tdelta}
\end{eqnarray}
$k_B$ being the Boltzmann constant.
On account of eqs. (4) and (\ref{100}), eq. (41) yields
\begin{equation}
\begin{array}{l}
0.5~\Xi^{-1}\left(\displaystyle{\frac{M_{DC}}{M_{\odot}}}\right)^{2/3}~{\rm K}
~\ut < ~T_{\Delta} ~\ut < 
16.2~ \times \\
f_{CL}~\Xi^{-2}  \left(\displaystyle{\frac{\rho(0)}{M_{\odot}{\rm pc^{-3}}}}
\right)
\left(\displaystyle{\frac{M_{\odot}}{M_{DC}}}\right)^{1/3}~{\rm K}~,
\label{42}
\end{array}
\end{equation}
which -- in the illustrative case $f_{CL} \simeq 0.5$,
$\rho(0) \simeq 3 \times 10^3
M_{\odot} {\rm pc^{-3}}$
and $M_{DC}\simeq 10^5~M_{\odot}$ -- entails in turn $5.3\times 10^3~
{\rm K}~\ut < T_{\Delta}\ut < ~1.3\times 10^4~{\rm K}$.
As a consequence of the increased temperature, the molecules within 
$\Delta$ will radiate, thereby reducing the excess energy in the
cloud. In order to see whether this mechanism actually prevents the
cloud from evaporating, we notice that the characteristic time needed
to accumulate the energy $\Delta E_c$ inside $\Delta$ is just the
traversal time $t_m$. Therefore, this energy will be totally radiated
away provided the cooling rate per molecule $\Lambda$ exceeds the critical
value $\Lambda_*$ given by the equilibrium condition
\begin{equation}
N_{\Delta} \Lambda_* t_m \simeq \Delta E_c~.  \label{37}
\end{equation}
Specifically, eq. (\ref{37}) yields
\begin{equation}
\Lambda_* \simeq 10^{-12}~ \Xi^{-1} \left(\frac{{\rm pc}}{r_m} \right)
\left(\frac{{\rm km}}{a_1}\right)^{3/2} ~{\rm erg~ s^{-1}~ mol^{-1}}~,
\label{38}
\end{equation}
on account of eqs. (15), (\ref{32}) and (\ref{NDelta}). 
Moreover, in the present case
in which most of the molecules are $H_2$ the explicit form of
$\Lambda$ is 
(see e.g. O'Dea et al. 1994, Neufeld et al. 1995)
\begin{equation}
\Lambda \simeq 3.8 \times 10^{-31}~\left(\frac{T_{\Delta}}{{\rm K}}
\right)^{2.9}~~{\rm erg~s^{-1}~H_2^{-1}}~, 
\label{27}
\end{equation}
which -- thanks to eq. (41) -- becomes
\begin{equation}
\Lambda \simeq 3.4 \times 10^{21} ~\Xi^{-2.9} \left(\frac{{\rm km}}
{a_1} \right)^{4.35}~ {\rm erg~s^{-1}~H_2^{-1}} \label{40}
\end{equation}
(notice that $\Lambda$ is almost independent of $M_{DC}$). Now,
from eqs. (\ref{38}) and (\ref{40}) it follows that the condition
$\Lambda \ut > \Lambda_*$ implies
\begin{equation}
a_1 \ut < ~6 \times 10^{11} ~\Xi^{-0.7} 
\left(\frac{r_m}{{\rm pc}}\right)^{0.35} ~{\rm km}~.
\label{410}
\end{equation}

Again, it is difficult to figure out the relevance of condition (\ref{410})
in general, but in the illustrative case of $M_{DC} \simeq 10^5 M_{\odot}$
it turns out to be abundantly met for hard primordial binaries.

Thus, we conclude that
the energy given up by primordial binary brown dwarfs and 
temporarily acquired by molecular clouds is efficiently radiated away,
so that the clouds do not get dissolved by frictional hardening.

As a final comment, we stress that our estimate for $T_{\Delta}$
should be understood as an upper bound, since thermal conductivity
has been neglected. In addition,  
the above analysis implicitly
relies upon the assumption $T_{\Delta} < 10^4$ K, which ensures the
survival of $H_2$. Actually, in spite of the fact that eq. (\ref{42})
entails that this may well not be the case, our conclusion remains  
nevertheless true. For, higher temperatures would lead to the depletion
of $H_2$, which correspondingly gets replaced by atomic and
possibly ionized hydrogen. As is well known, in either case the resulting
cooling rate would exceed the one for $H_2$, and so cooling would be
even more efficient than estimated above (B\"ohringer \& Hensler 1989).

\section{Conclusions}
We have shown that -- within the considered model -- 
the overwhelming majority of binary brown 
dwarfs is so hard today that their orbital radius is smaller than (roughly)
one-half of the Einstein radius for microlensing towards LMC, thereby
making them unresolvable in the microlensing experiments performed so far.
Still, we have seen that not too hard primordial binaries -- whose
initial orbital radius obeys condition (31) -- have an orbital radius
today which typically turns out to be about one order of magnitude
smaller than the above-mentioned Einstein radius. Therefore, we argue
that not too hard primordial binary brown dwarfs can be resolved in
future microlensing experiments with  a more accurate photometric
observation
\footnote{
See e.g. the ongoing experiments by the GMAN and PLANET collaborations
(Proceedings of the {\it Second International Workshop on Gravitational 
Microlensing Surveys}, Orsay, 1996).}, the signature being small
deviations from standard microlensing light curves (Dominik 1996).
Notice that such a procedure complements the detection strategy for
binaries suggested by Gaudi and Gould (1996). Finally, we point out
that the mechanism considered here can naturally account for an
average MACHO mass somewhat larger than $\simeq 0.1~M_{\odot}$.

\vspace*{2mm}

\noindent
{\bf ACKNOWLEDGEMENTS}

\noindent
We would like to thank the Editor for a helpful remark.
FDP has been partially supported by the Dr. Tomalla Foundation, 
GI by the Agenzia Spaziale Italiana and
MR by the Dipartimento di Fisica Nucleare
e Teorica, Universit\`a di Pavia, Pavia, Italy.


\begin{thebibliography}{ }
\bibitem[1996]{Adams}
Adams F.C., Laughlin G., 1996, ApJ, 468, 586
\bibitem{alcock}
Alcock C. et al., 1993, Nature 365, 621  
\bibitem{Alcock1}
Alcock C. et al., 1996, astro-ph 9606165 
\bibitem{Aubourg}
Aubourg E. et al., 1993, Nature, 365, 623 
\bibitem{bahcall}
Bahcall J. et al., 1994, ApJ, 435, L51 
\bibitem{bt}
Binney J., Tremaine S., 1987, {\it Galactic Dynamics}, Princeton 
University Press, Princeton
\bibitem{BH}
B\"ohringer H., Hensler G., 1989, A \& A 215, 147 
\bibitem{bhl}
Burrows A., Hubbard W.B., Lunine J.I., 1989, ApJ, 345, 939
\bibitem{carr} 
Carr B., 1994, Ann. Rev. Astron. Astrophys., 32, 531
\bibitem{csm}
Chabrier G., Segretain L., M\'era D., 1996, ApJ, 468, L21
\bibitem{dantona}
D'Antona F., 1987, ApJ, 320, 653
\bibitem{dijr}
De Paolis F., Ingrosso G., Jetzer Ph., Roncadelli M., 1995a, Phys. Rev. 
Lett., 74, 14 
\bibitem{depaolis}
De Paolis F., Ingrosso G., Jetzer Ph., Roncadelli M., 1995b, 
A \& A, 295, 567 
\bibitem{dijqr}
De Paolis F., Ingrosso G., Jetzer Ph., Qadir A., Roncadelli M., 1995c, 
A \& A, 299, 647 
\bibitem{vangelo}
De Paolis F., Ingrosso G., Jetzer Ph., Roncadelli M., 1996, 
Int. J. Mod. Phys., D5, 151 
\bibitem{dij} 
De Paolis F., Ingrosso G., Jetzer Ph., 1996, ApJ, 470, 493 
\bibitem{kn:Derujula} 
De R\'ujula A., Jetzer Ph., Mass\'o E., 1991, MNRAS, 250, 348
\bibitem{dominik}
Dominik M., 1996, Thesis, Dortmund University 
\bibitem{evans}
Evans N.W., 1996, astro-ph 9611161
\bibitem{ }
Fabian A.C., Nulsen P.E.J., 1994, MNRAS, 269, L33
\bibitem{ }
Fabian A.C., Pringle J.E., Rees M.J., 1975, MNRAS, 172, 15
\bibitem{fall}
Fall S.M., Rees M.J., 1985, ApJ, 298, 18 
\bibitem{fields}
Fields B.D., Mathews G., Schramm D.N., 1996, astro-ph 9603035
\bibitem{ggt}
Gates E.J., Gyuk G., Turner M., 1996, Phys. Rev., D53, 4138 
\bibitem{gg}
Gaudi B.S., Gould A., 1996, astro-ph 9606104
\bibitem{Gerhard} 
Gerhard O., Silk J., 1996, ApJ, 472, 34 
\bibitem{ }
Gibson B.K., Mould J.R., 1996, astro-ph 9612133
\bibitem{heggie}
Heggie D.C., 1975, MNRAS, 173, 729
\bibitem{hu}
Hu E.M. et al., 1994, Nature, 371, 493
\bibitem{kn:Jetzer} 
Jetzer Ph., 1996, Helv. Phys. Acta, 69, 179
\bibitem{kerins}
Kerins E.J., 1996, astro-ph 9610070
\bibitem{kerins2}
Kerins E.J., 1997, astro-ph 9704179
\bibitem{Lee}   
Lee H.M., Ostriker J.P., 1986, ApJ, 310, 176
\bibitem{kn:Maoz}
Maoz E., 1994, ApJ, 428, L5   
\bibitem{met}
Metcalf R.B., Silk J., 1996, ApJ 464, 218
\bibitem{}
Neufeld D.A., Lepp S., Melnick G.J., 1995, ApJS 100, 132
\bibitem{ }
Nulsen P.E.J., Fabian A.C., 1997, submitted
\bibitem{}
O'Dea C.P. et al., 1994, ApJ 422, 467 
\bibitem{kn:Paczynski} 
Paczy\'nski B., 1986, ApJ, 304, 1
\bibitem{kn:Paczynski} 
Paczy\'nski B., 1996, Ann. Rev. Astron. Astrophys., 34, 419
\bibitem{Persic}
Persic M., Salucci P., 1990, MNRAS, 247, 349
\bibitem{pt}
Press W.H., Teukolsky S.A, 1977, ApJ, 213, 183 
\bibitem[1996]{sackett}
Sackett P.D., 1997, ApJ, 483, 103
\bibitem{spitzer}
Spitzer L., 1987, {\it Dynamical Evolution of Globular Clusters},
Princeton University Press, Princeton
\bibitem{}
Spitzer L., Mathieu R.D., 1980, ApJ 241, 618
\bibitem{soc}
Statler T.S., Ostriker J.P., Cohn H.N., 1987,  ApJ, 316, 626 
\bibitem{tam}
Tamanaha C.M. et al., 1990, ApJ, 358, 164
\bibitem{trimble} Trimble V., 1987, Ann. Rev. Astron. Astrophys., 25, 425 
\bibitem{vanalbada}
van Albada T.S., Sancisi R. 1986, Phil. Trans. R. Soc.
London A 320, 447 
\bibitem{Zahn}
Zahn J.P., 1987, A \& A, 57, 383 

\end{thebibliography}
\end{document}